\documentstyle[11pt,rafringe,twoside]{article}

\def\edcomment#1{\iffalse\marginpar{\raggedright\sl#1\/}\else\relax\fi}
\reversemarginpar

\begin{document}
\title{Radio Galaxies and the Magnetization of the IGM}
\author{Gopal-Krishna}
\affil{National Centre for Radio Astrophysics/TIFR, Post Bag 3, Pune University
Campus, Pune 411 007, India 
}
\author{P. J. Wiita}
\affil{Department of Physics \& Astronomy,  MSC 8R0314, Georgia State University,
Atlanta GA, 30303-3088, USA 
}

\begin{abstract}

Observed radio galaxies had a much higher comoving density during
the `quasar era', at $z \sim 2-3$, but these sources are only detectable
for small fractions of their active lifetimes at such high $z$ due to  
expansion losses and increased inverse Compton losses against the cosmic
microwave background.  
Using recent models for the evolution of the size and luminosity of 
powerful double radio sources, as well as $\Lambda$CDM simulations of the 
cosmic web of baryonic material, we argue that during the quasar era a high 
volume fraction of this web was occupied by the lobes of double radio 
sources. They could have seeded the IGM with an average magnetic field 
approaching $10^{-8}$G. Further, these advancing overpressured lobes 
could compress the denser interstellar gas clouds of the galaxies engulfed by them
and thus trigger starbursts. This can probably account for 
much of the intense 
star-formation activity witnessed beyond $z \sim 1.5$. Also, the sweeping 
up of the ISM of the gas-rich galaxies by the rapidly advancing radio lobes 
may well be responsible for the widespread metal pollution of the IGM and 
proto-galaxies at high redshifts.

\end{abstract}

\section{Introduction}
The comoving space density of powerful radio galaxies (RGs) has  
declined between 100 and 1000
times from redshifts of 2--3 to the present epoch (e.g. Willott et al.
2001).  The star formation rate also peaked in roughly the same epoch,
and has declined substantially since then
(e.g. Archibald et al. 2001; Chary \& Elbaz 2001).  Severe
adiabatic and inverse Compton losses in radio lobes mean that most
old and large radio galaxies are very difficult to detect in surveys
and only the youngest can be seen at high redshifts
(e.g. Blundell et al. 1999, hereafter BRW).
Simulations of the growth of structures in cosmological models
have indicated that most of the matter that 
will form galaxies by the current epoch was in the form of filaments
that filled only a small portion of the universe at those redshifts
(e.g. Cen \& Ostriker 1999).  Together, these facts lead us to conclude that 
the formation of many of those  galaxies may have been triggered 
or accelerated by overpressured
radio lobes, which probably filled a substantial portion of the cosmic filaments then. 
Preliminary discussions of these ideas
are in Gopal-Krishna \& Wiita (2001, hereafter GKW) and 
Gopal-Krishna, Wiita \& Osterman (2003); more extensive calculations
are underway (Osterman et al.,
in preparation).
These extremely extended radio lobes are very likely to be responsible for 
seeding much of the intergalactic medium (IGM) with magnetic fields
(GKW; Kronberg et al. 2001; Furlanetto \& Loeb 2001).
In addition, they may make significant contributions to the spreading of
metals widely through the IGM and into protogalaxies.

\section{Radio luminosity function -- RLF}
In general, RLF studies are plagued by uncertainties resulting from incomplete
knowledge of the redshifts
of the radio sources; however, results based 
upon the 3CRR, 6CE and 7CRS surveys of different flux limits 
have the advantage of having 96\% of their redshifts known (Willott
et al. 2001).  In addition, their selection at low frequencies minimizes
the bias due to  relativistic beaming.

Powerful  (FR II type) radio sources are nearly
3 dex above the local RLF by $z \sim 2$, and their RLF varies little
out to the beginning of the quasar era at $z \sim 3$, 
while it appears to decline at higher $z$ 
(Willott et al. 2001).  Furthermore, the RLF for those redshifts is nearly
flat for over a decade in radio power above $P_{151} \ge 10^{25.5}$
W Hz$^{-1}$ sr$^{-1}$, which is where the FR II sources are most
numerous.  

Combining these results with the correction factor discussed 
in the following section,
we find that at $z = 2.5$ the actual proper density  of powerful 
radio sources lasting for an interval
$T$ is
\begin{equation}
\rho \simeq 4 \times 10^{-5}
(1+z)^3 T_5~{\rm Mpc}^{-3} (\Delta \log P_{151})^{-1},
\end{equation}
where $T_5 \equiv T/(5\times 10^8{\rm y})$.  
We then integrate
 over the
roughly 1.25 dex of the peak of the RLF.  Finally, we take into
account the fact that several generations of RGs will be born 
and die within the $\sim 2$ Gyr duration of the quasar era.  This leads
us to the total proper density, $\Phi$, of intrinsically powerful 
radio sources: $\Phi = 7.7 \times 10^{-3} ~{\rm Mpc}^{-3}$, which is
independent of the assumed value of $T$, and nearly independent of
cosmological model parameters (see GKW for details).

\section{Restricted visibility of radio galaxies}
Various recent models of RG evolution (Kaiser et al. 1997;
BRW; Manolakou \& Kirk 2002) agree that radio
flux declines dramatically with increasing source size (adiabatic
losses) and with $z$ (inverse Compton losses off the more intense
cosmic background
radiation).  Theoretical distributions of RG powers, sizes, 
redshifts and spectral indices can be nicely matched by
models that require most RGs to remain active for $T \simeq 5 \times 10^8$y
(see, also, Wang et al. 2002) and to have
a distribution of jet powers ($Q_0$) that goes as $\sim Q_0^{-2.6}$
(BRW).  X-ray observations indicate  that the density of
the matter through which the jets propagate declines with distance
roughly as
$n(r) = n_0(r/a_0)^{-\beta}$, with  $n_0 = 1.0 
\times 10^{-2}{\rm cm}^{-3}$, $a_0 = 10~$kpc, and
$\beta = 1.5$.  This leads to the total  linear size of the RG being given by
\begin{equation}
D(t) = 3.6 a_0 \left(\frac{t^3 Q_0}{a_0^5 m_p n_0}\right)^{1/(5-\beta)}.
\end{equation}

We find that in most RGs, particularly those
at $z > 2$, the central engines remain active for much longer times than 
those galaxies are detected in flux limited surveys, and therefore they 
should grow to very large linear sizes (typically $D(T) > 1$\,Mpc),
although detecting them would require extremely sensitive radio surveys
with redshifts.
Using the models of BRW we find that the visibility time, $\tau \propto Q_0^{1/2}$, 
and to properly estimate the actual number of RGs
from those detectable in flux-limited surveys, one must multiply
by a correction factor ($T/\tau$) of roughly 50 for powerful
RGs during the quasar era (GKW).  We are currently investigating how
this factor depends on the parameters defining the BRW
and Manolakou \& Kirk (2002) radio source evolution models.

The likelihood of `missing' faded radio lobes resulting from past activity is
underscored by the sophistication needed for detecting even the giant outer 
lobes of the nearby radio galaxy M87 (Owen et al. 2000). `Compact
double' radio sources, several of which are found to have faint, diffuse radio
lobes (e.g.  Schoenmakers et al. 1999; Baum et al. 1990; Stanghellini et al.
1990),
example of this difficulty.
Associating these faint structures with their true core, or even with each
other, would have been extremely difficult had these sources been imaged only a few hundred years ago when their present central radio components were not yet born.
Another indication of the weakening of radio lobes towards
higher redshifts is the apparent decrease in the radio-loud fraction of QSOs
with redshift, as recently inferred by Impey \& Petry (2001).

\section{Radio lobes pervade the relevant universe}
Recent numerical models of the evolution of  $\Lambda$CDM universes
indicate that at $z \sim 0$, roughly 70\% of baryons are in a cosmic web of
filaments of warm-hot gas and embedded galaxies and clusters that together
occupy only about 10\% of the volume of the universe  (e.g. Cen \& Ostriker 1999;
 Dav{\'e} et al. 2001).  But at $z \sim 2.5$ the growing network of filaments
comprised only about 20\% of the baryonic mass, and a quite small fraction, $\eta \simeq 0.03$,
 of the total volume.

It is certainly to be expected that the massive galaxies that harbor super-massive black 
holes large enough to
form RGs at early times would have typically formed in the densest portions
of those filaments, usually at their intersections. The radio lobes ejected 
from them would mostly remain within
the filaments, and since it is in this relatively small, `relevant universe',
that new galaxies formed out of denser gas clumps, we really only need to be concerned with what fraction of
this relevant universe the lobes permeated.  We find that the mean volume of
a radio source is $\langle V(T) \rangle \simeq 2.1 T_5^{18/7} ~{\rm Mpc}^{3}$, and 
thus,  the volume 
fraction of the  relevant universe which radio lobes born during
the quasar era cumulatively swept through is (GKW)
\begin{equation}
\zeta = \Phi ~\langle V (5 \times 10^8 {\rm yr}) 
\rangle ~(0.03/\eta)(5/R_T)^2 \simeq 0.5,
\end{equation}
where $R_T \sim 5$ is the typical length to width ratio of an RG; we have
conservatively ignored the volume filled by  the lobes of the
weaker FR I source population, 
which also  evolves rapidly (Snellen \& Best\ 2001).  
The energy
density injected by the lobes into the filaments is 
$u \simeq 2 \times 10^{-16}$\,J\,m$^{-3}$ for those same canonical parameters.  Because $\langle V(T) \rangle$ is
a sensitive function of $T$, if the typical RG lifetime is  $< 10^8$ yr, then
$\zeta < 0.01$ and $u < 9 \times 10^{-18}$\,J\,m$^{-3}$ (GKW).

\section{Extensive star formation at high$-z$ is triggered by radio lobes}
The discovery of the alignment effect between extended optical emission lines and
radio lobe directions (e.g. McCarthy et al. 1987; Chambers et al. 1987) quickly led to
calculations (e.g. Begelman \& Cioffi 1989; Rees 1989) that indicated that
star formation could be triggered by these expanding overpressured lobes.
This extended optical emission is produced to some degree by direct ionization
by the AGN and perhaps by shock heating, but there is often a significant
component best explained as due to radiation from young stars.
Recent  hydrodynamical simulations including cooling (Mellema et al. 2002) confirm 
that star formation is likely to occur through cloud fragmentation, cooling and compression.
{\sl HST} observations of high-$z$ RGs and associated
optical emission (e.g. Best et al. 1996; Bicknell et al. 2000) 
support this scenario.

 \vfill

To inquire if these lobes are capable of triggering
extensive star formation on larger scales we (GKW) used the models
of Falle (1991) and BRW to find that:
$p_{\rm lobe} \propto
t^{(-4-\beta)/(5-\beta)}$, but $D \propto t^{3/(5-\beta)}$, so
$p_{\rm lobe} \propto D^{(-4-\beta)/3}$. 
 The decline in the external pressure
is slower, $p_{\rm ext} \propto D^{-\beta}$, so
$p_{\rm lobe}/p_{\rm ext} \propto D^{(-4+2\beta)/3}$.  For
$\beta = 3/2$,  $p_{\rm lobe}/p_{\rm ext} \propto D^{-1/3}$. 
The values of $Q_0$, $\rho_0$ and $a_0$ 
appropriate for FR II sources produce overpressures at $D = 50$ kpc amounting
to factors of $10^2$--$10^4$, corresponding to Mach numbers of 10--100
 for the bow shock. 

\vfill

Thus, 
these  powerful RGs create lobes which 
typically remain rapidly expanding, with overpressures of factors
 $> 10$ (and Mach numbers above 3) out to distances
of well over 1 Mpc.  Supersonic expansion into a two-phase circumgalactic 
medium will compress many of the cooler gas clumps, rapidly reducing
the Jeans mass by factors of 10--100 and thereby 
triggering starbursts (Rees 1989; Mellema et al. 2002).
The clouds will then remain in substantially overpressured lobes
of low density, which can continue to produce extensive starbursts.

\vfill

Related situations have been considered by Evrard (1991) for
the infall of clouds into the hot ICM of clusters of galaxies, and by
Jog \& Das (1992; 1996) for the infall of a captured galaxy into the 
ISM of the captor galaxy. The latter authors examined
what happens to molecular clouds 
in a captured galaxy as they enter into the central portions of a 
larger galaxy where higher pressure gas is awaiting them.
Under many reasonable circumstances these clouds undergo radiative
shock compression; when the growth time for gravitational instabilities
in the shocked outer shell of the cloud becomes smaller than the
shock's crossing time (as it usually will), the fragmenting shell
should produce many stars in a rather efficient
fashion (Jog \& Das 1992; 1996).  

\vfill

It is thus  fair to argue that
many clumps of gas sitting in localized dark matter potential wells may
yield extensive starbursts, or even entire galaxies, after being
enveloped by an expanding radio lobe.  Hence, we predict an enhanced
2-point correlation function (and an even more enhanced 3-point correlation
function) between radio galaxies and newly formed galaxies during the quasar era.
Unfortunately, these correlations will be extremely difficult to measure, since the relevant
 radio sources will
have typically faded below detectability even while their expanding lobes
continue to have a major impact on the surrounding medium at substantial
distances from the AGN.
\newpage

\section{Radio lobes magnetize the IGM}

One exciting implication of this scenario is that RGs can inject a
substantial amount of magnetic energy into the IGM at $z \sim 2-3$.
Faraday rotation measurements of quasars provide a nominal upper bound on
intergalactic magnetic fields of $B_{IGM} < 10^{-9}$G (e.g. Kronberg et
al. 1999). However, if the magnetic field is 
preferentially distributed in the cosmic filaments where the 
relevant IGM is also concentrated, then fields within those filaments
ranging up to $\sim 10^{-6}$G are allowed by these observations
(Ryu et al. 1998). Kronberg et al. (1999) have argued that a substantial
fraction of the IGM may have been permeated by magnetized outflows
from stars in galaxies. The tentative
detection of magnetic fields in high-$z$ galaxies (e.g. Oren \& Wolfe 1995)
poses considerable difficulties for standard dynamo models,
since amplification in galactic disks requires many dynamical times
(e.g. Furlanetto \& Loeb 2001). 

The possibility of jets in radio galaxies
magnetizing the IGM has been examined previously, but those earlier
investigations concluded that either only minute
magnetization levels or insignificant volume coverage  would be attained
(e.g. Daly \& Loeb 1990; Rees 1994). In GKW we showed 
that during the quasar era, the permeation           
of the IGM by the expanded lobes of radio galaxies could have seeded the 
IGM with an average magnetic fields of $\approx 10^{-8}$G 
(recall that fields of $\approx 1\mu$G exist even inside the lobes of 
megaparsec RGs, e.g., Kronberg et al. 2001; Ishwara-Chandra \& Saikia\ 1999), 
which matches
the IGM field strengths inferred by Ryu et al. (1998) and by Furlanetto \& Loeb
(2001). The latter authors, advancing essentially orthogonal arguments to
ours (GKW), based on the evolution of isotropized magnetized bubbles fed by
quasars, have argued that the quasar population is capable
of polluting $\sim$ 10\% of the entire space with
magnetic fields. From independent arguments, Kronberg et al. (2001) have
concluded that the accretion energy released by radio-loud QSOs at $z \sim
2-3$ is adequate to magnetize the IGM to the level of its thermal energy,
provided the radio lobes can expand to fill up the IGM.

\section{Enriching new galaxies with metals}
Several authors have discussed the issue of `metal transport' from their
production sites, namely the ISM of galaxies, to the Mpc-scale IGM with a
mean density
$< 10^{-4}$ as high  (e.g. Gnedin 1998; Shchekinov 2002), 
and even farther into the voids (Theuns et al. 2002).   
Lyman-break galaxies at  $z > 3$ often have metalicities
around 0.1 solar and damped Lyman-$\alpha$ clouds have metalicities
 $\sim 10^{-2.5}$ solar (e.g. Steidel et al. 1999; Pettini et al. 2001).
 The difficulty is underscored further with the 
recent detection of O{\sc vi} absorption at $z \sim 2-3$ from under-dense 
regions ($\rho/\langle\rho\rangle < 1$) representing the true IGM (Schaye et al. 2000). 
  Supernova explosions in star-forming galaxies are found 
to fail by at least an order-of-magnitude to pollute the whole IGM to the 
metalicity levels observed within the available time (e.g. Gnedin \& 
Ostriker 1997; Ferrara et al. 2000).  
A possible alternative could be mergers of protogalaxies in the process 
of hierarchical clustering (Gnedin \& Ostriker 1997; Gnedin 1998). 

In  accordance with our picture of a radio lobe-filled universe at high-$z$, 
we propose here a potentially attractive new mechanism for large-scale 
metal transport: namely, the sweeping of the ISM of star-forming galaxies 
by the expanding giant radio lobes during the quasar era, or even earlier. 
The outflowing radio jets will drag along a significant
fraction of the gas in their host galaxy out with them, most of it compressed into
a shell along the bow shock outside the lobes, as illustrated by
numerical simulations of jets leaving a galaxy's ISM (e.g. Hooda \& Wiita 1998).  
This enriched gas can then be
spread over distances exceeding 1\,Mpc over the course of $\sim 10^8$ years.
Eventually, substantial portions of this gas can be mixed into the radio lobe,
but these Rayleigh-Taylor and Kelvin-Helmholtz types of instabilities grow 
relatively slowly under these conditions, and so it seems that
most of the enriched gas will comprise part of a shell that will also include swept up
ICM as the lobe continues to expand.  Note that
the morphology of the line-emitting gas surrounding the radio bubbles
of M87 suggests that much of this material has been excluded from the
cocoon (Bicknell \& Begelman 1996).
When this expanding gaseous shell interacts with denser clouds
in the ICM or IGM, not only will extensive star formation be triggered, but
this star-forming region will have incorporated a fraction of this swept-up
enriched gas.  In that much of it is likely to have remained in the bow-shock
region, the dilution will not be as severe as it would have been if 
the enriched gas were spread 
throughout the immense volume of the radio lobes.
An advantage of this mechanism is that heating of the denser gas is less of 
a problem than when the metals are conveyed by supernova driven winds.

In addition to the radio lobes contributing to ``metalization'' by
dragging along some of the enriched gas from their host galaxy,
it is worth considering their impact on other young galaxies which they
may envelop.  If these young galaxies have a multi-phase medium and are not
too dissimilar from local galaxies in this regard, then ram pressure stripping (Gunn \&
Gott 1972) produced by these lobes can be important.  Even if the average
density of the ICM has fallen as low as $10^{-5}$cm$^{-3}$ at $r = 1$\,Mpc,
as expected for $\beta = 1.5$,
the density of the shell will typically be compressed to several times this
value.  With the expansion velocity of the lobes remaining roughly $10^4$ km
s$^{-1}$, the ram pressure, $P_{\rm ram} \simeq \rho v^2 > 3 \times 10^{-11}$
 dyn cm$^{-2}$.
This pressure is adequate to remove most of the gas from a typical spiral
galaxy  (with $v_{\rm rot} \simeq 200$ km s$^{-1}$ and $R \simeq 10$ kpc; 
see Abadi et al. 1999; Fujita 2001), and is likely to be
even more effective in stripping the diffuse gas from smaller, 
recently forming galaxies
at $z > 2$, despite our lack of knowledge of the details of the properties
of that ISM.  However, colder, denser clumps of gas within those 
galaxies will probably
not be pushed out (cf. Mellema et al. 2002); rather, the arriving 
lobe may trigger yet more star formation
in those regions.  

Finally, we note that individual AGN may go through several
generations of nuclear activity that yield radio jets and lobes.
The first such episode could trigger extensive star formation, or
even new galaxy formation, in relatively nearby clouds.  Any subsequent 
lobes hitting that
newly formed galaxy a few hundred Myr later could sweep out most of the 
enriched gas it had already
produced, thereby propagating these newer metals into yet more distant regions, in
the fashion discussed in the last paragraph.  These metals could, in turn,
contribute to the seeding of additional clouds which are triggered to collapse by this
second (or subsequent) period of activity.

\section{Conclusions} 
Even though the  local universe is very sparsely populated by powerful radio sources, 
several large factors conspire
 to make them remarkably
important for galaxy formation during the quasar era.   
First, their comoving density was roughly
1000 times higher at $\sim 2 < z < \sim 3$.  Second, only a small fraction (roughly
two percent) of the powerful sources present during that period are detected
in the surveys used to produce the RLFs, because of severe inverse 
Compton and adiabatic losses; these unseen, old radio lobes
fill very large volumes.  Third, the fraction
of the volume of the universe occupied by the material
during the quasar epoch that would finally condense into clusters of
galaxies was only a few percent, so these lobes only had to permeate this `relevant
universe' rather than the entire universe.  In that the best models of RG evolution
indicate that the lobes  are overpressured and supersonically expanding into
the relevant universe, the scenario that has many massive starbursts, and even 
many galaxies, formed in this fashion at high redshifts, is quite plausible. 

Another key result is that radio galaxies were likely to have been capable of
seeding the IGM with magnetic fields of the appropriate strength.  It
is very encouraging that two other independent and quite different arguments
lead to similar conclusions (Furlanetto \& Loeb 2001; Kronberg et al. 2001).
Also, the sweeping up of the ISM of the galaxies and star-forming clouds
by the expanding lobes of RGs suggests a natural way to spread metals 
produced in the first stellar generations over large volumes.  Detailed 
investigations of many aspects of this scenario are clearly warranted.

\acknowledgements
GK thanks Prof. Marshall Cohen for the invitation to this memorable meeting 
held in Ken's honor, and NRAO for partial travel support.
PJW is grateful for support from RPE funds at GSU and for 
continuing hospitality at the Department of Astrophysical
Sciences at Princeton.

\end{document}